\documentclass[12pt]{iopart}

\usepackage{graphicx} \usepackage{bm} \usepackage{dcolumn}

\begin{document}
\title[Zitterbewegung in CNT created by laser pulses]{Zitterbewegung of electrons in carbon nanotubes created by laser pulses}
\date{\today}
\author{Tomasz M Rusin$^1$ \ and Wlodek Zawadzki$^2$}
\address{$^1$ Orange Customer Service sp. z o. o., Al. Jerozolimskie, 02-326 Warsaw, Poland\\
         $^2$ Institute of Physics, Polish Academy of Sciences, 02-668 Warsaw, Poland}
\ead{Tomasz.Rusin@orange.com}

\pacs{72.80.Vp, 42.50.-p, 41.75.Jv, 52.38.-r}
\submitto{\JPC}

\begin{abstract}
We describe a possibility of creating non-stationary electron wave packets in zigzag carbon nanotubes (CNT) illuminated
by short laser pulses. After disappearance of the pulse the packet experiences the trembling motion (Zitterbewegung, ZB).
The band structure of CNT is calculated using the tight-binding approximation generalized
for the presence of radiation. Employing realistic pulse and CNT parameters we obtain the ZB oscillations
with interband frequencies corresponding to specific pairs of energy bands.
A choice of optimal parameters is presented in order to observe the phenomenon
of ZB experimentally. The use of Gaussian wave packets to trigger the electron Zitterbewegung, as used in the
literature, is critically reexamined.
\end{abstract}

\maketitle

\section{Introduction}

The phenomenon of Zitterbewegung (ZB, trembling motion) was devised by Schrodinger in 1930~\cite{Schroedinger1930}.
Schrodinger observed that, in the~$4\times 4$ Dirac Hamiltonian for relativistic electron in a vacuum,
the velocity operator~$\hat{v}_i = \partial {\hat H} /\partial {\hat p}_i$ does not commute with the Hamiltonian,
so that, even in the absence of external fields, the electron velocity is not a constant of the motion. Schrodinger
calculated velocity and position operators as functions of time and showed that they contain, in addition to the expected
classical terms, also quickly oscillating terms, which he called Zitterbewegung. This quantum result is quite
unexpected, as it goes beyond the Newton first law of motion. Since Schrodinger's pioneering work the phenomenon
of ZB has become a subject of numerous theoretical papers and some controversies. It has been recognized that,
mathematically speaking, ZB is a result of interference between the positive- and negative-energy states of the
Dirac Hamiltonian, see e.g.~\cite{BjorkenBook}. Lock~\cite{Lock1979} in his important paper
showed that, if one represents the electron as a wave packet then, as a result of the Riemann-Lesbegues
lemma, the ZB oscillations decay in time. The use of wave packets satisfies the well-known requirement of the quantum theory
stating that an operator alone does not represent physical reality. The latter is given by an average value of the operator
taken over a quantum state. This difficulty is present in the original treatment of Schrodinger's, who obtained his results
in terms of operators.

Beginning from 1970 the Zitterbewegung was proposed also for electrons in solids where the velocities are not
relativistic~\cite{Lurie1970,Cannata1990,Vonsovskii1990,ZawadzkiHMF}.
The decisive feature is that the electron energy spectrum consists of at least two bands, so that
the interference of upper and lower energy states can take place. A real surge of works dealing with ZB in solids
and other periodic systems began in 2005, when the papers by Zawadzki~\cite{Zawadzki2005} and
Schliemann {\it et al.}~\cite{Schliemann2005} dealing with semiconductors appeared. As we already mentioned,
representing an electron by a wave packet makes theoretical treatment much more physical, although it introduces
a decay of ZB oscillations which depends on packet's width. However, there exists another important parameter
defining the initial conditions. Since the ZB phenomenon is a result of interference between upper and lower
energy states, the initial state must also be defined by upper and lower initial components. In handbooks on
relativistic quantum mechanics~\cite{BjorkenBook,GreinerBook,WachterBook} four-component packets are usually assumed
without specifying their values. For semiconductors, commonly assumed packets have the Gaussian shape with one
non-vanishing component~\cite{Schliemann2005,Rusin2007b,Zhang2008a,Clark2008,Demikhovskii2010}
\begin{equation} \label{Intr_Gauss}
 \langle{\bm k}|F\rangle \propto \left(\begin{array}{c}1 \\ 0 \end{array} \right) \exp\left[-\frac{1}{2}d^2({\bm k}-{\bm k}_0)^2\right].
\end{equation}
For this choice, as we show in Appendix~A, one always needs an initial momentum~$\hbar{\bm k}_0$ in one
direction to have ZB oscillations in the perpendicular direction. On the other hand,
as shown by Gerritsma {\it et al.}~\cite{Gerritsma2010}
in their proof-of-principle simulation of the one-dimensional Dirac equation with the use of trapped ions
interacting with laser beams, a Gaussian wave packet was created with two non-vanishing
components and the observed Zitterbewegung had the same direction as the initial momentum.
A review of ZB in various periodic systems was given by the present authors in~\cite{Zawadzki2011}.

It follows from the above description that the choice of initial wave packet, i.e. its shape, initial momentum
and components, is decisive for the resulting properties of ZB. As to the common choice of packets
indicated in~(\ref{Intr_Gauss}), which requires the initial value of~$\hbar{\bm k}_0$, it is not
clear how to create this momentum. If one uses light to create the electron packet,
the initial wave vector~${\bm k}_0$ will be small since photons do not carry much momentum.
If one uses acoustic phonons to trigger the ZB motion, the initial energy will be small. One could
use simultaneously energy and momentum conservation laws in photon excitations between bands, but then
one would deal with packets depending on the modulus~$|{\bm k}_0|$, i.e.
with a circle (or sphere) of momenta in different directions around the zero value. Also, a narrow
light pulse creates a relatively large energy uncertainty which results in a wide spread
of initial momenta.

In view of the above difficulties and also to make the project more realistic experimentally,
we propose here not to assume anything {\it a piori} about the electron wave packet, but to
determine it as a result of a realistic laser pulse. Then we propose to use it for a calculation of
ZB oscillations. A preliminary effort to determine and then use the electron
wave packet created by a light pulse was carried out in a perturbative way in~\cite{Rusin2009}.
We consider carbon nanotubes (CNT) as a suitable electronic system, see~\cite{Rusin2007b,Zawadzki2006}.
We also tried to apply our procedure to monolayer graphene, but it turned out that graphene
is not particulary suitable for our purpose. This aspect is discussed in Appendix~B.

The paper is organized as follows. In Section~2 we outline the theory of packet creation and
ZB oscillations in carbon nanotubes with the use of an ultra-shot laser pulse.
In Section~3 we present the results of calculations, in Section~4 we
discuss the obtained results. The paper is concluded by a summary.
In Appendices we discuss an influence of packet components on ZB in general,
the ZB motion in graphene for the electron wave packet created by a laser pulse,
and the validity of an approximate form of the vector potential.

\section{Zitterbewegung in carbon nanotubes}

\begin{table}
 \begin{tabular}{|c|l|l|}
   \hline

   $a$    & lattice constant & 2.46~\AA \\
   $k$    & wave vector & $k \in [-\pi/a\sqrt{3},\pi/a\sqrt{3}]$ \\
   $t_{AB}$ & matrix element for atoms placed & \\
            &  in~$A$ and~$B$ points & 3.03 eV \\
   ${\bm C}_h$ & chiral vector~${\bm C}_h=(N,0)$ & $(9,0)$ \\
   \hline
   $\omega_L$ & central laser frequency & 4.5~fs$^{-1}$ \\
   $\tau$ & pulse length & 4.5~fs \\
   $E_0$  & field intensity & $4\times 10^9$~V/m \\
   $t_E$  & time of pulse termination & 19~fs \\
   \hline
   $\omega_6$ & $2E_{m=6}/\hbar$ & 0 \\
   $\omega_7$ & $2E_{m=7}/\hbar$ & 4.91~fs$^{-1}$ \\
   $\omega_5$ & $2E_{m=5}/\hbar$ & 6.02~fs$^{-1}$ \\
   $\omega_8$ & $2E_{m=8}/\hbar$ & 8.11~fs$^{-1}$ \\
   \hline
 \end{tabular}
 \caption{Parameters for a zigzag carbon nanotube after~\cite{Saito1992,SaitoBook} (first box),
          laser pulse parameters after~\cite{Wirth2011} (second box),
          frequencies corresponding to the gap energies obtained from the tight-binding theory (third box).}
\end{table}

Here we present the theory of ZB oscillations in carbon nanotubes after illumination of the system
by a short laser pulse. At the beginning we summarize the tight-binding
approach to the energy bands in CNT
and then introduce the vector potential of laser field to the formalism.
Following Saito {\it et al.}~\cite{Saito1992,SaitoBook} we
consider monolayer graphene in which carbon atoms are placed in two nonequivalent
points of the hexagonal lattice, called traditionally the~$A$ and~$B$ points.
Each atom placed in the~$A$ point is surrounded by three atoms placed in the~$B$ points, whose relative positions
are:~${\bm R}_1=a/\sqrt{3}(1,0)$,~${\bm R}_2=a/\sqrt{3}(-1/2,\sqrt{3/2})$ and~${\bm R}_3=a/\sqrt{3}(-1/2,-\sqrt{3/2})$,
where~$a=2.46$~\AA\ is the length of carbon-carbon bond.
Within the usual tight-binding approximation one expands the Bloch function of the electron into a linear
combination of~$\phi_A$ and~$\phi_B$ atomic functions in~$A$ and~$B$ points.
The matrix elements of the periodic Hamiltonian~${\cal H}$ between atoms in two~$A$ or two~$B$ points
vanish, while the matrix element of~${\cal H}$ between the atomic functions in~$A$ and~$B$ points is
\begin{equation} \label{ML_HAB1}
 {\cal H}_{AB}= \sum_{j=1}^3 t_j({\bm R}_j) e^{i{\bm k}\cdot {\bm R}_j},
\end{equation}
where~$t_j({\bm R}_j)$ are transfer integrals between the atom in~$A$ point and the atom in~$j$th~$B$ point
\begin{equation} \label{ML_tRj}
 t_j({\bm R}_j) = \langle \phi_A({\bm r}) |{\cal H}| \phi_{Bj}({\bm r}-{\bm R}_j) \rangle.
\end{equation}
In the absence of fields there is:~$t_j({\bm R}_j)=t_{AB}$ for all~$j$,
and the tight-binding Hamiltonian for the electron in graphene is
\begin{equation} \label{ML_hH}
\hat{H} = t_{AB} \left(\begin{array}{cc} 0 & {\cal H}_{AB}^* \\ {\cal H}_{AB} & 0 \end{array} \right),
\end{equation}
where
\begin{equation} \label{ML_HAB2}
 {\cal H}_{AB}= e^{ik_xa/\sqrt{3}} + 2e^{-ik_xa/(2\sqrt{3})}\cos\left(\frac{k_ya}{2}\right).
\end{equation}

A nanotube is obtained from a graphene sheet by rolling it into a cylinder.
As a result of folding, one joins lattice points connected
by the chiral vector~${\bm C}_h=n_1 {\bm a}_1 + n_2{\bm a}_2\equiv (n_1,n_2)$,
where~${\bm a}_1= a(3/2,\sqrt{3}/2)$ and~${\bm a}_2=a(3/2,-\sqrt{3}/2)$
are lattice vectors and~$n_1$,~$n_2$ are integers.
After wrapping, the wave vector~$k_y$ parallel to~${\bm C}_h$ becomes quantized,
while the~$k_x$ vector parallel to tube's
axis remains continuous, so that nanotube's band structure presents a set of one-dimensional energy bands.
Here we consider a zigzag nanotube characterized by the chiral vector~${\bm C}_h=(N,0)$, and the quantizing
condition is:~$k_y=(2\pi/a)(m/N)$ with~$m=1 \ldots 2N$~\cite{Saito1992,SaitoBook}.
The tight-binding Hamiltonian for CNT can be obtained from~(\ref{ML_hH})
by replacing~${\cal H}_{AB}$ by
\begin{equation} \label{CNT_HAB2}
 {\cal H}_{AB}^T= e^{ika/\sqrt{3}} + 2e^{-ika/(2\sqrt{3})}\cos\left(\frac{m\pi}{N}\right),
\end{equation}
in which~$k \in [-\pi/a\sqrt{3},\pi/a\sqrt{3}]$, where we write~$k_x=k$.
The energy bands in CNT obtained from~(\ref{ML_hH}) and~(\ref{CNT_HAB2}) are
\begin{equation}
 E_{k,m}= \pm t_{AB}|{\cal H}_{AB}^T|.
\end{equation}
For given~$k$, the energy~$E_{k,m}$ forms~$4N$ energy bands
symmetric with respect to~$E=0$, labeled by two quantum numbers:~$m=1 \ldots 2N$ and the energy sign~$\epsilon=\pm 1$.
In figure~1 we plot energy bands in the vicinity of~$E=0$ for zigzag CNT with~$N=9$. Note that for~$J=1 \ldots 8$
each pair of energy bands with~$m=9-J$ is degenerate with the pair having~$m=9+J$.
The bands with~$m=9$ and~$m=18$ are not degenerate. For other
properties of energy bands in CNT see~\cite{SaitoBook}.

\begin{figure}
\includegraphics[width=8.0cm,height=8.0cm]{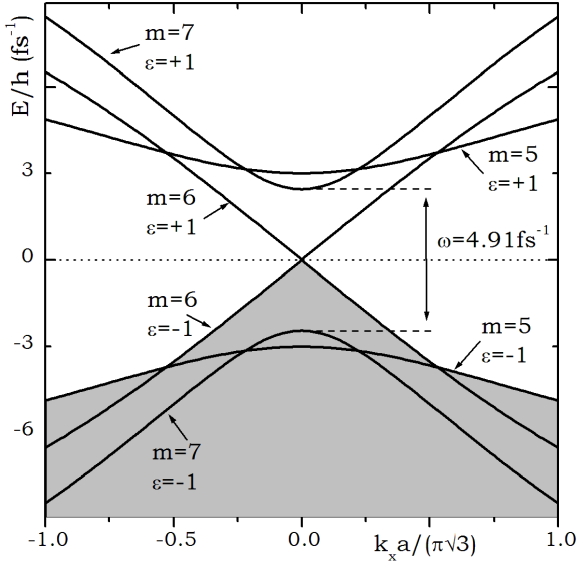}
\caption{Schematic plot of energy bands close to~$E=0$ in zigzag CNT for parameters listed in
         table~1, see~\cite{SaitoBook}.} \label{Fig1}
\end{figure}

Now we want to describe an effect of illumination of CNT by laser light.
We consider a single laser pulse whose electric field oscillates in the~$x$ direction.
Within the electric dipole approximation the electric field is
\begin{equation} \label{Pulse_Et}
 E(t) = E_0\exp\left(-b\frac{t'^2}{\tau^2}\right)\sin(\omega_Lt'),
\end{equation}
where~$t'=t-t_0$,~$E_0$ is the field intensity,~$\tau$ is the pulse duration,~$\omega_L$ is central laser
frequency,~$b=2\ln2\simeq 1.386$ and~$t_0=2.5\tau$ is the time shift of the pulse center.
Pulses characterized by parameters listed in table~1
were created experimentally, see~\cite{Wirth2011}.
We introduce light by the vector potential~${\bm A}={\bm A}(t)$
and the scalar potential~$\phi=0$. Then~${\bm E}=-\partial {\bm A}(t)/\partial t$
and~${\bm B}={\bm \nabla} {\bm \times} {\bm A}(t)=0$.
By choosing~${\bm A}(t)=[A(t),0,0]$ we have approximately
\begin{equation} \label{Pulse_At}
 A(t) \simeq \frac{E_0}{\omega_L}\exp\left(-b\frac{t'^2}{\tau^2}\right)\cos(\omega_Lt').
\end{equation}
This approximation is not crucial to our problem but the analytical form of~$A(t)$
simplifies numerical description of the electron motion. For further discussion see Appendix~C.

In order to introduce the vector potential into the tight-binding Hamiltonian~(\ref{ML_hH})
we employ the method proposed by Graf and Vogl~\cite{Vogl1995}.
Following this approach we replace in~(\ref{ML_HAB1}) each~$t_j({\bm R}_j)$ by its potential-dependent counterpart~$T_j({\bm R}_j)$
\begin{equation} \label{ML_TRj}
 T_j({\bm R}_j) = t_j({\bm R}_j) \exp\left\{-\frac{ie}{2\hbar}{\bm R}_j\cdot[{\bm A}({\bm 0},t)+{\bm A}({\bm R}_j,t)]\right\}.
\end{equation}
We do not modify the in-site energies~${\cal H}_{AA}={\cal H}_{BB}=0$ because the scalar potential is zero.
Since~${\bm A}(t)$ in~(\ref{Pulse_At}) does not depend
on~${\bm R}_j$, we have
\begin{equation} \label{ML_TRj2}
 T_j({\bm R}_j) = t_{AB} \exp\left\{-\frac{ie}{\hbar}{\bm R}_j \cdot {\bm A}(t) \right\},
\end{equation}
and one obtains
\begin{eqnarray} \label{ML_HABt}
 \tilde{H}_{AB} &=& t_{AB} \sum_{j=1}^3 e^{i {\bm k}\cdot {\bm R}_j }e^{-i(e/\hbar){\bm R}_j \cdot {\bm A}(t)} \nonumber \\
                &=& t_{AB} \sum_{j=1}^3 e^{i {\bm q}(t)\cdot {\bm R}_j},
\end{eqnarray}
where~${\bm q}(t)= {\bm k} -(e/\hbar){\bm A}(t)$ is the generalized quasi-momentum.
The final result of the above approximations resembles the usual ``minimal coupling'' substitution for
the free-electron case in the presence of an electric field.

Thus, as a result of illumination by light, we deal with the time-dependent electron Hamiltonian for our problem.
The electron wave function~$\Psi(t)=(\Psi_1(t),\Psi_2(t))$ in CNT evolves in time
according to the Schrodinger equation
\begin{equation} \label{CNT_hHt}
i\hbar\frac{d}{dt} \left(\begin{array}{c} \Psi_1(t) \\ \Psi_2(t)\end{array} \right) =
 t_{AB} \left(\begin{array}{cc} 0 & h_0(t)^* \\ h_0(t) & 0 \end{array} \right)
 \left(\begin{array}{c} \Psi_1(t) \\ \Psi_2(t)\end{array} \right),
\end{equation}
where
\begin{equation} \label{CNT_h0}
 h_0(t)= e^{iq(t)a/\sqrt{3}} + 2e^{-iq(t)a/(2\sqrt{3})}\cos\left(\frac{m\pi}{N}\right),
\end{equation}
is the time-dependent counterpart of~$ {\cal H}_{AB}^T$ in~(\ref{CNT_HAB2}).
For each instant of time the time-dependent Hamiltonian~$\hat{H}(t)$, defined in the
right-hand-side of~(\ref{CNT_hHt}),
has two eigenvectors~$w_1(t)$ and~$w_2(t)$
corresponding to the positive and negative eigenenergies~$\Lambda(t)=\pm t_{AB}|h_0(t)|$, respectively.
There is
\begin{equation} \label{CNT_w1}
 w_1(t)=\frac{1}{\sqrt{2}|h_0(t)|} \left(\begin{array}{c} |h_0(t)| \\ h_0(t)\end{array} \right),
\end{equation}
and
\begin{equation} \label{CNT_w2}
 w_2(t)=\frac{1}{\sqrt{2}|h_0(t)|} \left(\begin{array}{c} -h_0^*(t) \\ |h_0(t)|\end{array} \right).
\end{equation}

We assume that that in the absence of fields the Fermi level is at~$E=0$,
so that all bands with negative energies are occupied and those with positive energies are empty.
For~$t=0$, when the electric field is not turned on yet, the initial condition for~$\Psi(t)$ is:~$\Psi(0)=w_2(0)$
for every~$k$ and~$m$. For given~$k$ and~$m$, the electron velocity in the~$x$ direction averaged
over the wave function~$\Psi(t)$, is~\cite{Rusin2013a}
\begin{equation} \label{CNT_vxkxmy}
 \langle v^{k,m}(t) \rangle=\left\langle \Psi(t)\left|\frac{\partial \hat{H}(t)}{\hbar\partial k}\right|\Psi(t)\right\rangle.
\end{equation}
The total average electron velocity integrated over~$k$ and summed over subbands is
\begin{equation} \label{CNT_vx}
 \langle v(t)\rangle=\frac{1}{2\pi}\sum_{m=1}^{2N} \int_{-k_{max}}^{k_{max}} \langle v^{k,m}(t) \rangle dk,
\end{equation}
where~$k_{max}=\pi/(a\sqrt{3})$. We calculate~$\langle v(t)\rangle$ numerically in a few steps.
First, we select~$2M+1$ values of~$k$ in the range~$|k| \le k_{max}$,
where~$M=200$. Next, for given~$k$ and~$m$ we solve~(\ref{CNT_hHt}) using the fifth order Runge-Kutta method.
The obtained wave packets have two non-zero components.
Then, we calculate the average electron velocity for each of~$m$ bands
\begin{equation} \label{CNT_vxmy}
 \langle v^{m}(t)\rangle = \frac{1}{2\pi}\int_{-k_{max}}^{k_{max}} \langle v^{k,m}(t) \rangle dk,
\end{equation}
and finally compute the total average velocity~$\langle v(t)\rangle$ in~(\ref{CNT_vx}).

We also calculate the probability~${\cal P}^-(t)$ of finding the electron in states
with negative time-dependent energy~$\Lambda(t)=-t_{AB}|h_0(t)|$
\begin{equation} \label{CNT_pm_kxmx}
 {\cal P}^-(t)= \frac{1}{2\pi}\sum_{m=1}^{2N} \int_{-k_{max}}^{k_{max}} {\cal P}^-_{k,m}(t) dk,
\end{equation}
where, for given~$k$ and~$m$
\begin{equation} \label{CNT_pm}
 {\cal P}^-_{k,m}(t) = |\langle\Psi(t)|w_2(t)\rangle|^2,
\end{equation}
is the probability distribution of finding electron in states with negative energy.
To find the meaning of~${\cal P}^-_{k,m}(t)$ let us expand the function~$\Psi(t)$ in terms of
eigenstates~$w_1(t)$ and~$w_2(t)$ of the Hamiltonian~(\ref{CNT_hHt}), see~(\ref{CNT_w1}) and~(\ref{CNT_w2}),
\begin{equation} \label{CNT_a1a2}
 \Psi(t) = a_1(t)w_1(t) + a_2(t)w_2(t).
\end{equation}
Then~${\cal P}^+_{k,m}(t)=|a_1(t)|^2$ and~${\cal P}^-_{k,m}(t)=|a_2(t)|^2$.
For~$t \geq t_E \simeq 19$~fs, when the pulse disappears,~${\cal P}^+_{k,m}(t)$ measures
the population of the upper energy level
excited by the laser pulse, and~${\cal P}^-_{k,m}(t)$ gives the same for the lower energy level.
Note that for~$t=0$ there is~${\cal P}^+_{k,m}(t)=0$ since all electrons occupy valence states only.
The necessary condition for the appearance of ZB is that both~${\cal P}^{\pm}_{k,m}(t)$ do not vanish,
which is achieved by the excitation of electron due to the laser pulse.
The results for~$\langle v(t)\rangle$ and~${\cal P}^-(t)$ are presented in the following.

\section{Results}

\begin{figure}
\includegraphics[width=8.0cm,height=8.0cm]{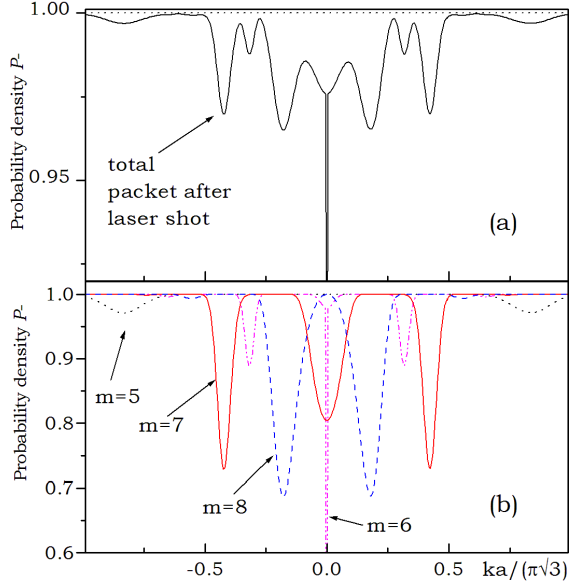}
\caption{(a) Calculated probability density of negative energy states (not normalized) for total wave packet in CNT
         created by the laser pulse at~$t_E=19$~fs at which the electric field of the pulse vanishes.
         (b) The same decomposed into four sub-packets created by the laser pulse in~$m=5,6,7,8$ bands.
         CNT and pulse parameters are listed in table~1.} \label{Fig2}
\end{figure}

\begin{figure}
\includegraphics[width=8.0cm,height=8.0cm]{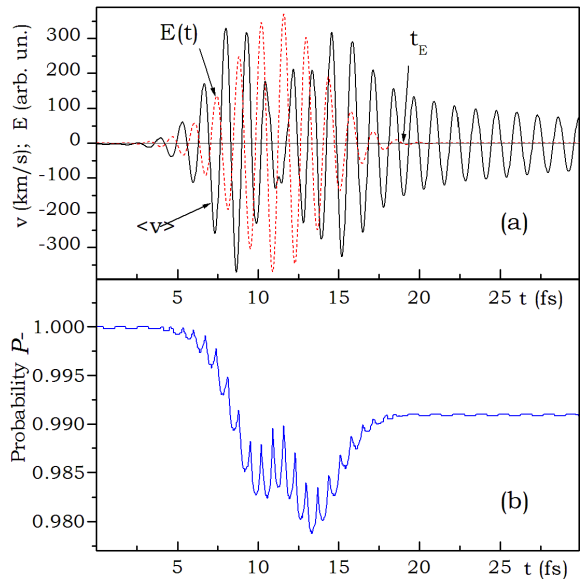}
\caption{(a) Calculated average packet velocity versus time calculated for CNT and laser pulse parameters listed in table~1.
         Solid line: packet velocity, dashed line: electric field
         of the laser pulse (in arbitrary units). Arrow indicates time~$t_E$ at which the electric field of the pulse vanishes.
         For~$t > t_E$ the amplitude of oscillations decays as~$t^{-1}$.
         (b) Normalized probability~${\cal P}^-$ of finding the electron in states with negative energy versus time.
        } \label{Fig3}
\end{figure}

In figure~2 we plot the probability density for negative-energy component of the wave packet
versus~$k$, see~(\ref{CNT_a1a2}), as calculated for the pulse and CNT parameters given in table~1.
Figure~2(a) shows the probability distribution~${\cal P}^-_{k}(t_E)$ at~$t_E=19$~fs,
for which the laser pulse terminates and the packet oscillates according to the field-free
Hamiltonian given in~(\ref{ML_hH}) and~(\ref{CNT_HAB2}). The packet created by the laser pulse consists
mainly of negative-energy states. Since there
is:~${\cal P}^+_{k}(t_E)=1-{\cal P}^-_{k}(t_E)$, the several minima of~${\cal P}^-_{k}(t_E)$
give rise to the positive-energy component of the packet.
The contribution of these states does not exceed a few percent of
the probability density. For~$k=0$ the probability of finding the electron in the
negative-energy state is zero (not shown in the figure), since the nanotube of~$N=9$ includes one pair of bands
(with~$m=6$) having the vanishing energy gap.

In figure~2(b) we show the calculated probability densities~${\cal P}^-_{k,m}(t_E)$
for four pairs of bands listed in table~1. For all pairs of bands, except the pair with~$m=6$, the
probability densities of negative energy states exceed 70\% and for each pair of bands they form a few bell-like minima.
It is seen that, for the pair of bands with~$m=7$
(i.e. those having the lowest non-vanishing energy gap),
the probability density~${\cal P}^-_{k}(t_E)$~has a bell-like minimum near~$k=0$,
which does not occur for other pairs of bands. As mentioned above,
the bell-like minimum of the negative-energy states corresponds
to the analogous maximum of the positive-energy states.
Thus, it is expected that the interference between positive and negative energy states
of the sub-packet created by the pair of bands with~$m=7$ may lead to the ZB oscillations
analogous to those described by the two-band~${\bm k} \cdot {\bm p}$ model in~\cite{Rusin2007b}.
In the following we show that this is in fact what happens.

In figure~3(a) we plot the calculated average electron velocity
versus time for the material and pulse
parameters listed in table~1. The solid line shows velocity oscillations of the electron packet excited
by the pulse, while the dashed line shows the electric field of the pulse (in arbitrary units).
At the initial time~$t=0$ both the electron velocity and electric field are zero.
Then, within the first~$19$~fs, the amplitude if electron velocity grows and decreases similarly
but not identically to the field amplitude.
After~$t_E \simeq 19$~fs the electric field of the pulse disappears,
but there persist oscillations slowly decaying in time.
These oscillations resemble the ZB oscillations in CNT described in~\cite{Rusin2007b} and they have a
similar character, slowly decaying as~$t^{-\alpha}$ with~$\alpha \simeq 1$.
In figure~3(b) we show the calculated time-dependent
probability of negative-energy component of the packet, see~(\ref{CNT_pm}).
Initially, for~$t=0$, the electron occupies only the valence states and there is~${\cal P}^-(0)=1$.
For~$t \ge t_E$, after the pulse disappears, the final electron state has a nonzero
admixture of states with positive energies which is a necessary condition for
the appearance of ZB oscillations~\cite{GreinerBook,WachterBook}.

We calculated the Fourier transform of~$\langle v(t)\rangle$ and the power
spectrum~$I_{\omega}\propto |\langle v(\omega)\rangle|^2$ for~$t\ge t_E$.
The results are shown in figure~4 by the solid line.
The intensity has the maximum for~$\omega\simeq 4.9$~fs$^{-1}$, which
is close to the interband frequency~$\omega_Z\simeq 4.91$~fs$^{-1}$ between the pair of
bands corresponding to~$m=7$, see figure~1. This means that the average packet velocity oscillates mostly with
the interband frequency~$\omega_Z$. The dashed line shows power spectrum of
the laser pulse given in~(\ref{Pulse_Et}) centered around~$\omega_L=4.5$~fs$^{-1}$ and
having the width~$\sigma\simeq 0.45$~fs$^{-1}$, which corresponds to pulse duration~$\tau=4.5$~fs.
Since the ZB frequency~$\omega_Z$ occurs within the pulse spectrum, it is possible to excite ZB oscillations.
The results presented in figures~3 and~4 confirm the possibility
of experimental creation of the wave packet which, for~$t\ge t_E\simeq 19$~fs,
(i.e. after the pulse termination) oscillates
with the interband ZB frequency~$\omega_Z\simeq 4.9$~fs$^{-1}$.

\begin{figure}
\includegraphics[width=8.0cm,height=8.0cm]{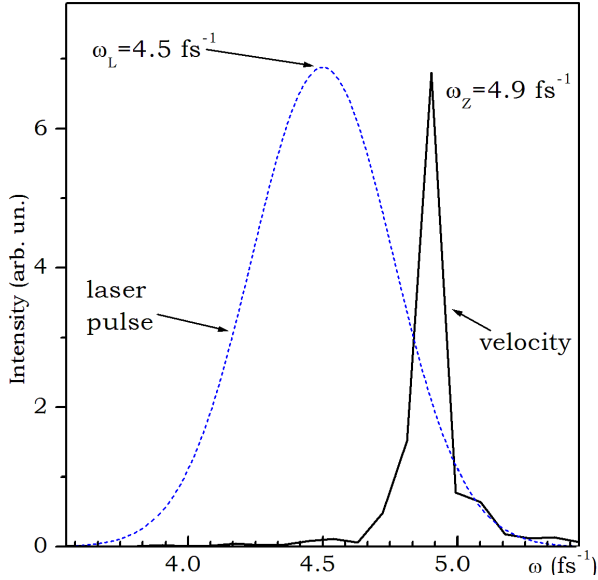}
\caption{Solid line: power spectrum of the Fourier transform of electron
         velocity~$I_{\omega} \propto |\langle v_x(\omega)\rangle|^2$,
         calculated for~$t\ge t_E$, versus frequency. The maximum of intensity is for the
         frequency~$\omega_{max} \simeq \omega_Z= 4.9$~fs$^{-1}$ (see figure~1). Dashed line:
         power spectrum of laser pulse (see equation~(\ref{Pulse_Et})) with
         central frequency~$\omega_L=4.5$~fs$^{-1}$.} \label{Fig4}
\end{figure}

The Fourier analysis of the electron velocity for~$t\le 19$~fs shows that, also in the presence
of electric field, the electron oscillates with the interband frequency~$\omega_Z \neq \omega_L$.
Although the packet oscillations have the ZB frequency both for~$t\le 19$~fs and~$t> 19$~fs,
we interpret only the motion at~$t> 19$~fs, in the absence of the laser pulse,
as an unmistakable manifestation of the ZB phenomenon. Figures~3 and~4 show the main results of our work.

\begin{figure}
\includegraphics[width=8.0cm,height=8.0cm]{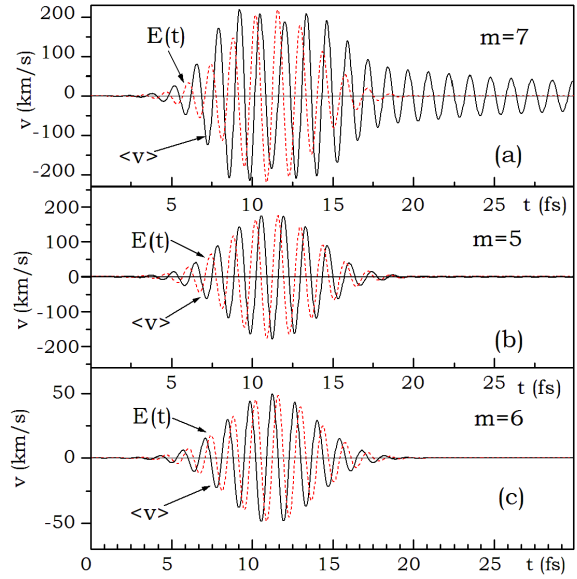}
\caption{Average packet velocity versus time, as calculated separately for three pairs of bands plotted in figure~1.
         Pulse parameters:~$\omega_L=4.5$~fs$^{-1}$,~$\tau=4.5$~fs and~$E_0=4 \times 10^9$~V/m.
         Material parameters are listed in table~1.} \label{Fig5}
\end{figure}

\begin{figure}
\includegraphics[width=8.0cm,height=8.0cm]{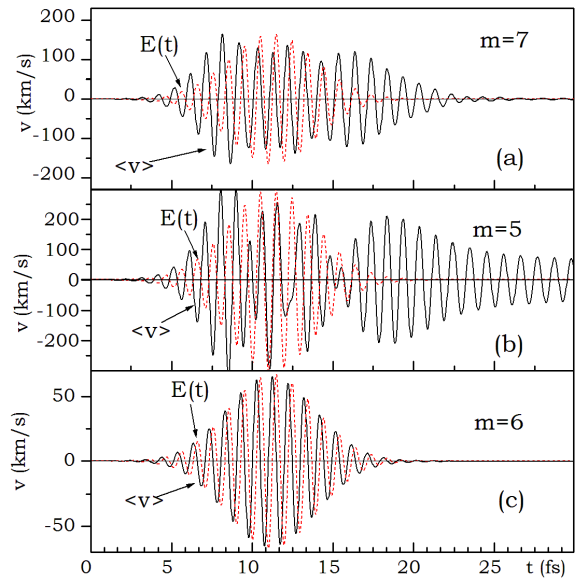}
\caption{The same as in figure~5 but with
        pulse parameters:~$\omega_L=6.4$~fs$^{-1}$,~$\tau=4.5$~fs and~$E_0=8 \times 10^9$~V/m.} \label{Fig6}
\end{figure}

Next, we analyze the electron motion in more detail.
Figure~5 shows separate velocities~$\langle v^{m}(t)\rangle$ resulting from the interference between
three pairs of energy bands, respectively.
For~$t\ge t_E$ the oscillations disappear for all pairs of bands except those for~$m=7$.
The latter have the interband ZB frequency~$\omega_Z\simeq 4.91$~fs$^{-1}$. This means that, for~$t\ge t_E$,
the ZB oscillations shown in figure~3(a) are caused by the interference of states in upper and lower~$m=7$ bands.
They are similar
to those found within the~${\bf k}\cdot {\bm p}$ theory in~\cite{Rusin2007b}. In figure~6 we show
the calculated average velocity~$\langle v^{m}(t)\rangle$ for
the central laser frequency~$\omega_L$ close to the interband frequency~$\omega_Z=6.02$~fs$^{-1}$.
The latter corresponds to the pair of bands with~$m=5$.
In this case, for~$t\ge t_E$ the ZB oscillations for the pair of bands with~$m=7$ quickly vanish,
but the oscillations for the bands with~$m=5$ resemble
those in figure~5(a), which means that the ZB motion results from the~$m=5$ bands.
Results shown in figures~5 and~6 indicate that ZB oscillations are caused by pairs of bands separated by the gap
close to the energy of laser light.
This means that, in principle if the gap between bands falls within the range of power spectrum of the laser pulse,
these bands can contribute to ZB motion. This, however, should not be understood as single-photon resonances.
The calculations shown in figures~5(c) and~6(c) strongly suggest that the
pair of linear bands with~$m=6$, for which the energy gap vanishes,
{\it does not} contribute to the ZB motion after the termination of the laser pulse.
We discuss this feature in Appendix~B.

\begin{figure}
\includegraphics[width=8.0cm,height=8.0cm]{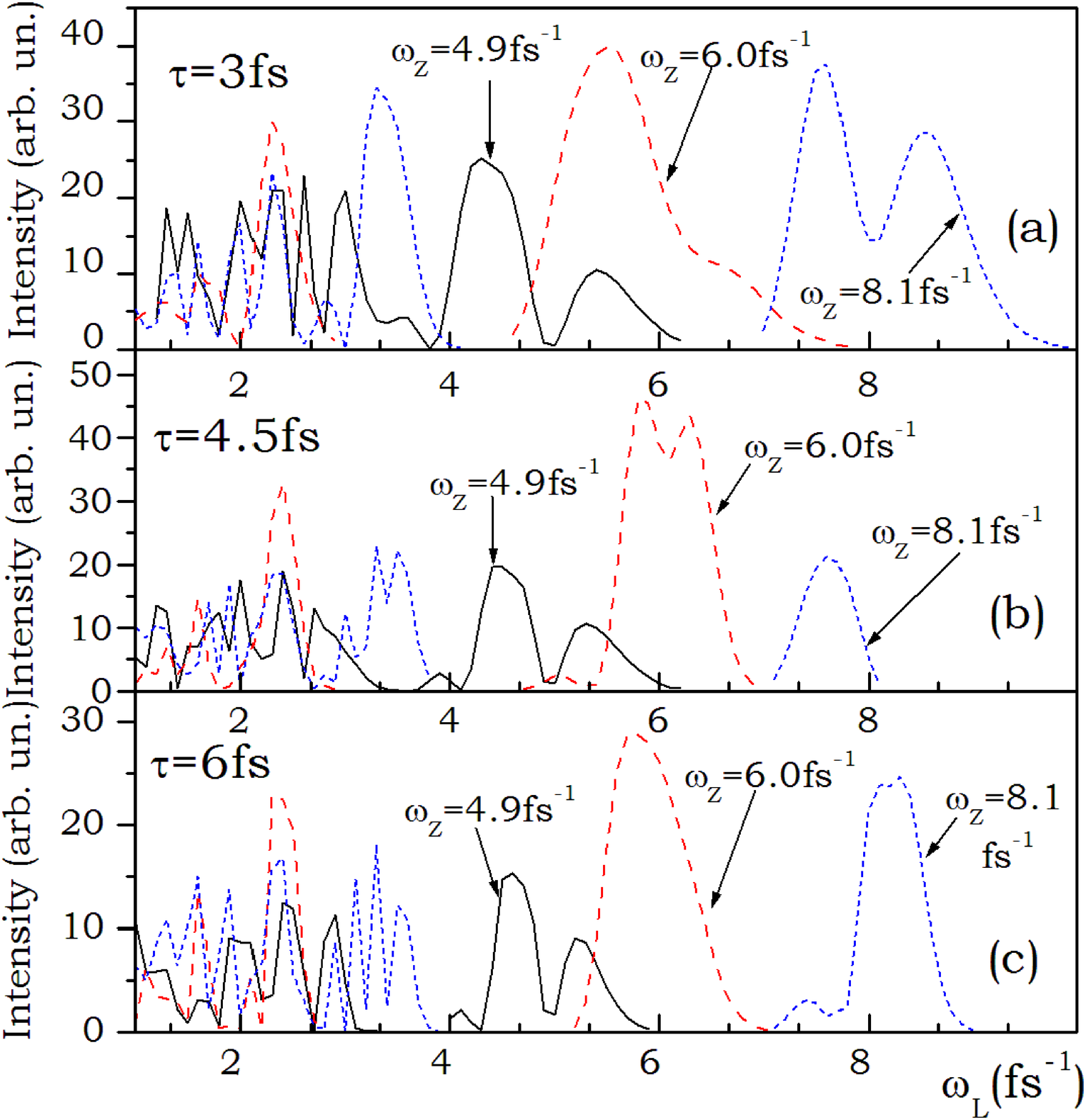}
\caption{Power spectrum of three Fourier transforms of the electron velocity~$|\langle v_m(\omega)\rangle|^2$ vs.
         central laser frequency~$\omega_L$ for three values of~$\tau$.
         Field intensity is~$E_0=8 \times 10^9$~V/m. Solid lines: intensities of velocity component oscillating with
         interband frequency~$\omega_{7}=4.91$~fs$^{-1}$;
         long-dashed lines: the same for interband frequency~$\omega_{5}=6.01$~fs$^{-1}$; dotted lines: the same for
         interband frequency~$\omega_{8}=8.11$~fs$^{-1}$.} \label{Fig7}
\end{figure}

Finally, we analyze the intensities of main components of the average packet velocity for other pulse parameters.
For three values of pulse duration~$\tau$ we calculated the power spectrum of the Fourier transform of
packet velocity~$|\langle v_m(\omega)\rangle|^2$ as functions of laser frequency~$\omega_L$.
For each combination of~$\omega_L$ and~$\tau$ we calculated the packet
velocity~$\langle v(t) \rangle$, found the appropriate cut-off time~$t_E$ and calculated
the Fourier transform of~$\langle v(t) \rangle$ for~$t\ge t_E$ and its modulus squared.
In figure~7 we plot our results for the pairs of bands with~$m=5$,~$m=7$, and~$m=8$.
We identify two qualitatively different regimes of the velocity oscillations.
The first regime occurs for~$\omega_L \ge 4$~fs$^{-1}$, where the electron motion consists of
one or two interband frequencies close to~$\omega_L$.
The second regime occurs for~$\omega_L \le 4$~fs$^{-1}$, i.e.
for the frequencies below the energy gap for bands with~$m=7$.
In this regime the photon energy is {\it smaller} that the lowest non-vanishing
energy gap in CNT, but the large amplitude of the electric field excites some electrons from
the valence to the conduction bands.
This statement should be understood qualitatively. The excitement in this regime has no one-one correspondence
to the resonant frequencies.
As a result, in the second regime it is possible to obtain the ZB
oscillations with {\it a few} interband frequencies for sufficiently large electric field intensities.
In our opinion the ZB motion can be observed more readily in the first regime of~$\omega_L$.

In more general terms, one can ask the question whether the wave packet created in real space by the laser
pulse is sufficiently localized, so that one can talk about its trembling motion. We calculated its
variance:~$\langle\hat{x}^2\rangle - \langle\hat{x}\rangle^2$ and it turns out that
it behaves like that of an equivalent Gaussian packet slowly spreading in time. This means that one can
legitimately consider the trembling motion of such a packet.

\section{Discussion}

It is more convenient to measure experimentally the effect of ZB when there is only one ZB frequency.
An existence of two or more frequencies, as indicated in figure~7,
may be inconvenient since then one must calculate the Fourier transforms of observed oscillations.
In this paper we concentrate on packet's velocity rather than position, because the velocity is
more closely related to observable quantities, like radiation or current. An average position
can be obtained from the velocity by the time
integral:~$\langle x(t) \rangle = \int_{-\infty}^t \langle v(t') \rangle dt'$.

It should be mentioned that we analyzed previously the ZB motion in CNT using the~${\bm k}\cdot {\bm p}$
method assuming the Gaussian shapes of electron wave packets as given in~(\ref{Intr_Gauss}),
see~\cite{Rusin2007b}. In this case,
averaging over the packet presents no difficulties because the Gaussian function automatically
limits the range of integration. The use of~${\bm k}\cdot {\bm p}$ method for the packets
created by laser pulses, which are not strongly concentrated in the~${\bm k}$ space, see figure~2,
creates problems of averaging because the final results depend on the range of integration.
For this reason we used in this paper the tight-binding method, so that the integration
is automatically limited to the first Brillouin zone. The second important reason for using the
tight-binding calculations for CNT is that they give a few realistic
energy bands which play different roles for various laser frequencies~$\omega_L$.
The tight-binding model for CNT, as considered in this paper, includes only the interactions between
valence~$p_z$ orbitals of the nearest-neighbour atoms, while other orbitals, i.e.~$s$,~$p_x$ and~$p_y$,
are neglected. As pointed out in~\cite{Reich2002}, more advanced models for CNT
give similar energy band structures to that employed here.

We consider zigzag CNTs with the chiral vector~${\bm C}_h =(9,0)$.
Such CNTs were widely analyzed in the literature, both experimentally and theoretically~\cite{SaitoBook}.
It is of interest to consider other possible CNTs having, for example, other chiral vectors~${\bm C}_h$.
Taking~$N=7$ allows one to observe ZB with the interband
frequency~$\omega_Z=2.28$~fs$^{-1}$, which can be excited by a red laser, as discussed in~\cite{Wirth2011},
or with the Ti:Sapphire laser. However, the obtained ZB oscillations in this case are much
weaker than those shown in figure~3. We also calculated the electron motion for wider
nanotubes:~$N=17$,~$N=25$ and~$N=53$.
By increasing~$N$, it is possible to reduce the gap between positive and negative energy bands even
to 0.32~fs$^{-1}$, i.e. to the far-infrared region.
However, illumination of such nanotubes by short laser pulses causes excitations of electrons
in many bands, so the resulting electron motion consists of many interband frequencies.
For~$N=53$ the packet motion for~$t\ge t_E$ resembles chaotic noise
and it is only after calculating the Fourier transforms that one can identify various interband components.
Therefore, it seems impractical to observe the ZB motion in wider CNTs.

In addition to the zigzag nanotubes described by the chiral vector~${\bm C}_h=(N,0)$ there
exist two other types of nanotubes: armchair CNTs, characterized by a
chiral vector~${\bm C}_h=(N,N)$, and chiral CNTs having a chiral vector~${\bm C}_h=(N_1,N_2)$
with integer~$N_1,N_2$. In the present paper we chose the zigzag CNT because in this case the energy gaps are located at
the~$\Gamma$ point of the Brillouin zone (at~$k=0$), while for armchair
or chiral CNTs the energy gaps occur in low-symmetry points of the Brillouin zone~\cite{Saito1992}.
This feature complicates analysis of ZB, but the ZB oscillations should
also exist in armchair or chiral nanotubes.

In our model we consider completely occupied valence bands and completely empty conduction bands
which assumes that the system is at low temperatures. This approach can be justified by the results
presented in figures~5 and~6, showing that the ZB oscillations are caused by the pair of~$m=7$
energy bands separated by the wide energy gap~$\hbar\omega_Z \simeq 3.2$~eV.
Since the linear energy bands with~$m=6$ do not contribute to the ZB motion,
one may safely neglect the impact of nonzero temperatures.
We disregard both the many-body and scattering effects.
This is an approximation, since the scattering as well as radiative recombination
processes will cause a decay of the non-stationary states.
After the excitation, packet's energy may be larger
than the energy of initial state. This excess energy will be radiated through
recombination processes which may be different
from the radiation related to oscillating electron dipoles due to the ZB motion.
As pointed out before~\cite{Rusin2009,JacksonBook}, the oscillating electron packet emits
dipole radiation proportional to the average packet's acceleration.
The dipole radiation has the same frequencies as~$\langle v(t) \rangle$, so its measurement
may confirm the existence of packet motion with the ZB frequencies.

Excitation of ZB oscillations requires sufficiently short laser pulses properly tuned to the
interband frequency. It has been recently possible to produce laser pulses with the relative
carrier-to-envelope phase~(CEP)~\cite{Kres2006}. For such pulses the electric field has
the form:~$E(t)={\cal E}(t)\sin(\omega_Lt+\phi_{CEP})$, where~${\cal E}(t)$ is the Gaussian
envelope and~$\phi_{CEP}$ is the carrier-to envelope-phase. The results shown in Section~3 correspond
to~$\phi_{CEP}=0$, see~(\ref{Pulse_Et}). We also performed calculations for~$\phi_{CEP}=\pi/2$
and the results are qualitatively similar to those presented above.
We concentrate on experimentally available pulses having the intensity~$E_0=4\times 10^{9}$~V/m~\cite{Wirth2011}.
We also performed calculations for~$E_0$ changing from~$0.01\times 10^{9}$~V/m to~$64\times 10^{9}$~V/m,
and other parameters listed in table~1. For low fields:~$E_0 \ll 1\times 10^{9}$~V/m,
the electron motion is weakly perturbed by pulse's field and the resulting amplitude of ZB motion is small.
For larger fields, close to~$1\times 10^{9}$~V/m, we obtain clear ZB oscillations
whose amplitude depends on field intensity. The optimum field for packet creation
is~$E\simeq 8\times 10^{9}$~V/m, see figures~6 and~7, and for such a field the excited packet
consists of 95\% of states with negative energies and 5\% of states with positive energies.
For still larger fields,~$E\ge 25 \times 10^{9}$~V/m, the results become ambiguous
since a strong electric field excites electrons from many bands in a way similar to
that discussed in figure~7.

As to the gauge aspects, it is convenient to use the vector gauge since then the
Schrodinger equation~(\ref{CNT_hHt}) represents a set of ordinary differential equations for~$\Psi(t)$.
In the scalar gauge:~${\bm A}=0$,~$\phi=-eE(t)x$, the corresponding Schrodinger equation
presents a set of partial differential equations in time and position variables,
which is more difficult to solve.

If we treated our problem using the ${\bm k} \cdot {\bm p}$ formalism, the interband matrix elements
of momentum, necessary to determine the band structure, appear also in the interband optical transitions.
They would then be responsible for the resulting electron wave packet created by the laser pulse.
However, in the tight-binding formalism, when we calculate the electron packet solving
exactly the equations of motion by numerical procedure, the resulting packet is calculated
without approximations and it is probably quite realistic.

The average packet velocity calculated in figures~3,~5 and~6 are closely related to the interband
polarization induced by the laser pulse:~$P(t)=-e\int_{-\infty}^t \langle v(t')\rangle dt'$.
In fact, the ZB oscillations are proportional to the time-derivative of interband polarization.
An advanced method to calculate the latter quantity in CNT is the so-called
CNT Bloch equations (CBE)~\cite{Hirtschulz2008,Malic2008,Waxenegger2011}, which are modification
of semiconductor Bloch equations~\cite{HaugBook,KlingshirnBook} for carbon nanotubes. Thus one
can also use CBE to calculate the ZB oscillations. The formalism of CBE allows one to take into
account decoherence and relaxation processes. In a recent paper we took into account the above
features in the description of ZB using the density matrix formalism~\cite{Rusin2014a}. It turned out
that the decoherence causes an exponential decay of ZB oscillations governed by the decoherence
time~$T_2$. The experimental time~$T_2$ was estimated to be~$130$~fs~\cite{Voisin2003}, which is
much larger than the scale of ZB oscillations in figures~3,~5 and~6, so that the dephasing processes
do not modify qualitatively our results. The approach used here corresponds to
the free-particle version of CBE and it does not take into account excitonic effects
as well as the renormalization of the band energies by the many-electron interactions~\cite{KlingshirnBook}.
As follows from our analysis, the ZB oscillations depend on the energy gap between the pair of
interfering bands, so that they will be somewhat affected by the excitonic effects and gap
renormalization, but we do not expect any major modifications of our description caused by these effects.
We mention that, in order to introduce the SBE one often uses the so-called rotating wave
approximation (RWA)~\cite{Hirtschulz2008,Malic2008,Waxenegger2011}. In our paper we do not use this
approximation since our results are obtained numerically. However, in the regimes shown in figures~3,~5 and~6
the RWA would be justified since the differences between the central laser frequency and ZB frequency
are around~10\%-20\%.

Now we briefly mention numerical aspects of our calculations.
The results presented in figures~3 --~6 are obtained for a grid of~$2M+1$ points
with~$M=200$ and for~$2N$ energy bands, which means that~(\ref{CNT_hHt}) is solved~7,218 times.
We treat~$M$ as an accuracy-control parameter and the reported value of~$M$ ensures
the proper accuracy of~$k$ integration [see~(\ref{CNT_vx})].
The results in figure~7 require calculations of ZB for~90 values of~$\omega_L$ and for three values of~$\tau$,
so that~(\ref{CNT_hHt}) has to be solved~1,948,860 times. The norm of the wave
function~$\Psi(t)$ serves as a control-parameter of the numerical solution of~(\ref{CNT_hHt}).
We adjusted the time-step and other program parameters to ensure the norm of the wave function with
the relative error below~$10^{-6}$ for all instants of time.

The basic question arises: are the one-component shifted Gaussian wave packets, as used in the
literature and indicated in~(\ref{Intr_Gauss}), adequate to describe the phenomenon of electron
Zitterbewegung in solids? On the one hand, they correctly give ZB oscillations with the interband
frequency and, for example, the ZB motion for~$t\ge t_E$ calculated for a realistic wave packet,
as shown in figure~3(a), is similar to that calculated with the use of Gaussian packets,
see figure~3 of~\cite{Rusin2007b}.
This can be interpreted as an indication that ZB is a robust and basic phenomenon, not very sensitive
to the details of the employed model. However, there are two important objections to using the
shifted Gaussian packets. The first is, that such packets are almost impossible to create, mostly due to
their large built-in initial momentum. The second objection is that sometimes such packets are
``too good to be true'', resulting for example in very nice ZB oscillations in graphene,
see e.g. figure~2 in~\cite{Rusin2007b},
while the realistic wave packet created by the laser pulse gives almost no ZB effect, see our figure~8.
We conclude that one must be very cautious when using convenient but not realistic
shifted Gaussian packets for the description of Zitterbewegung.

Finally, it should be mentioned that a calculation similar to ours can be done for electrons in superlattices
with finite energy gaps between 2D subbands. Technically, such a calculation is somewhat more complicated
than that for the 1D motion in CNT since one should in addition integrate over the 2D electron wave vectors.

\section{Summary}

We describe the phenomenon of electron Zitterbewegung with the use of non-stationary electron
wave packets created by laser pulses in zigzag carbon nanotubes. The band structure of CNT is calculated
with the use of a tight-binding approximation generalized for the presence of radiation.
The laser light must be roughly tuned to the energy gap between proper pairs of energy bands in CNT.
It is shown that, after the laser pulse terminates, the electron experiences ZB oscillations
with the interband frequency. The pair of linear energy bands, separated by the vanishing energy gap,
does not contribute to the ZB effect. The influence of initial packet components on the resulting
ZB motion is analyzed and it is shown that electron wave packets commonly used in the theoretical
literature are convenient for calculations but not realistic.

\appendix

\section{Components of wave packets}

Here we briefly analyze, as a matter of example, the motion of an electron in monolayer graphene prepared
in the form of a packet with two nonzero components
\begin{equation}
\langle {\bm k}| f\rangle = \left(\begin{array}{c} a_{\bm k} \\ b_{\bm k} \end{array} \right),
\end{equation}
where~$a_{\bm k}$ and~$b_{\bm k}$ are two real functions of~${\bm k}$
normalized to:~$\int (|a_{\bm k}|^2 +|b_{\bm k}|^2) d^2{\bm k}=(2\pi)^2$.
In the~${\bm k}\cdot {\bm p}$ approach the Hamiltonian for electron at the~$K$ point of the Brillouin zone
is~\cite{Wallace1947,Novoselov2004}
\begin{equation} \label{AppA_HM}
\hat{H}_M = u\hbar \left(\begin{array}{cc} 0 & k_x-ik_y \\ k_x+ik_y & 0 \end{array} \right),
\end{equation}
where~$u\simeq 10^6$ m/s. The average packet velocity in the~$x$ direction
is~$\langle v_x(t) \rangle = \langle f|e^{i\hat{H}_Mt/\hbar}(\partial \hat{H}_M/\partial \hbar k_x )e^{-i\hat{H}_Mt/\hbar}|f\rangle$,
see~\cite{Rusin2007b}. By using the identity:~$e^{i\hat{H}_Mt/\hbar}=\cos(ukt)+i(\hat{H}_M/u\hbar k)\sin(ukt)$,
where~$k=\sqrt{k_x^2+k_y^2}$, we find after some manipulations
\begin{eqnarray} \label{ApplA_vx_ab}
\langle v_x(t) \rangle &=& \frac{2u}{(2\pi)^2}\int a_{\bm k}b_{\bm k} \frac{k_y^2}{k^2}\cos(2ukt) d^2{\bm k}
                           +\frac{2u}{(2\pi)^2}\int a_{\bm k}b_{\bm k} \frac{k_x^2}{k^2} d^2{\bm k} \nonumber \\
                       &+& \frac{u}{(2\pi)^2}\int (a_{\bm k}^2-b_{\bm k}^2) \frac{k_y}{k} \sin(2ukt) d^2{\bm k}.
\end{eqnarray}
The above integrals depend on the parities of~$a_{\bm k}$ and~$b_{\bm k}$ functions in the~$(k_x,k_y)$ space.
For a one-component packet with~$a_{\bm k}\neq0$ and~$b_{\bm k}=0$, commonly used in previous works,
the first two integrals in~(\ref{ApplA_vx_ab}) vanish and the remaining integral is an odd function of~$k_y$.
This integral vanishes for~$a_{\bm k}^2$ being an even function in~$k_y$, but is nonzero
for~$a_{\bm k}^2$ having a non-vanishing odd part. The common form of~$a_{\bm k}^2$ used in the
literature is~\cite{Schliemann2005}
\begin{equation} \label{AppA_ak}
 a_{\bm k}^2\propto \exp\left[-d^2k_x^2-d^2(k_y-k_{0y})^2\right].
\end{equation}
Then the odd-part of~$a_{\bm k}^2$ is
\begin{eqnarray} \label{AppA_ak_odd}
a_{\bm k}^{2\ (odd)}&=& \frac{1}{2}\left(a_{\bm k}^2 - a_{-\bm k}^2 \right) \nonumber \\
                    &\propto & e^{-d^2k_x^2} \left[e^{-d^2(k_y-k_{0y})^2} - e^{-d^2(k_y+k_{0y})^2}\right]. \ \ \
\end{eqnarray}
For~$k_{0y}=0$ there is also:~$a_{\bm k}^{(odd)}=0$, then the last integral in~(\ref{ApplA_vx_ab}) vanishes
and the average velocity is zero. However, for~$k_{0y} \neq 0$ there is~$a_{\bm k}^{2\ (odd)} \neq 0$, so that the last integral
in~(\ref{ApplA_vx_ab}) remains finite and it leads to decaying ZB oscillations~\cite{Schliemann2005}. Therefore,
for one-component Gaussian packets, the nonzero value of~$k_{0y}$ is a necessary condition of the ZB motion
in the~$x$ direction. This was concluded in many papers.

Let us now consider packets with two nonzero components.
For simplicity we take~$b_{\bm k}=a_{\bm k}$, with~$a_{\bm k}$
given in~(\ref{AppA_ak}), and set~$k_{0y}=0$. Then the last integral in~(\ref{ApplA_vx_ab}) vanishes identically, but the
first two integrals remain finite because they are even functions of~$k_x$ and~$k_y$. Then the first integral describes
ZB oscillations, while the second gives a rectilinear motion. This simple example shows that
various choices of packet components lead to qualitatively different results for the average velocity.

As to the wave packet created by a laser shot, see (\ref{CNT_hHt}), numerical calculations show
that after pulse termination the packet consists of~$1\%$ of states having positive energies and~$99\%$ of states
having negative energies. This packet has two non-zero components
\begin{equation}
\Psi( k)_{|t=t_E} = \left(\begin{array}{c}
 a_k^{(1)} + i a_k^{(2)}\\ b_k^{(1)} + i b_k^{(2)}\end{array} \right),
\end{equation}
where~$a_k^{(j)}$ and~$b_k^{(j)}$ are oscillating functions of~$k$. Since the created wave packet has
two non-vanishing components and it consists of states having both positive and negative energies one can expect
existence of the ZB oscillations. This is indeed the case, as shown above.

\section{ZB in monolayer graphene}

\begin{figure}
\includegraphics[width=8.0cm,height=8.0cm]{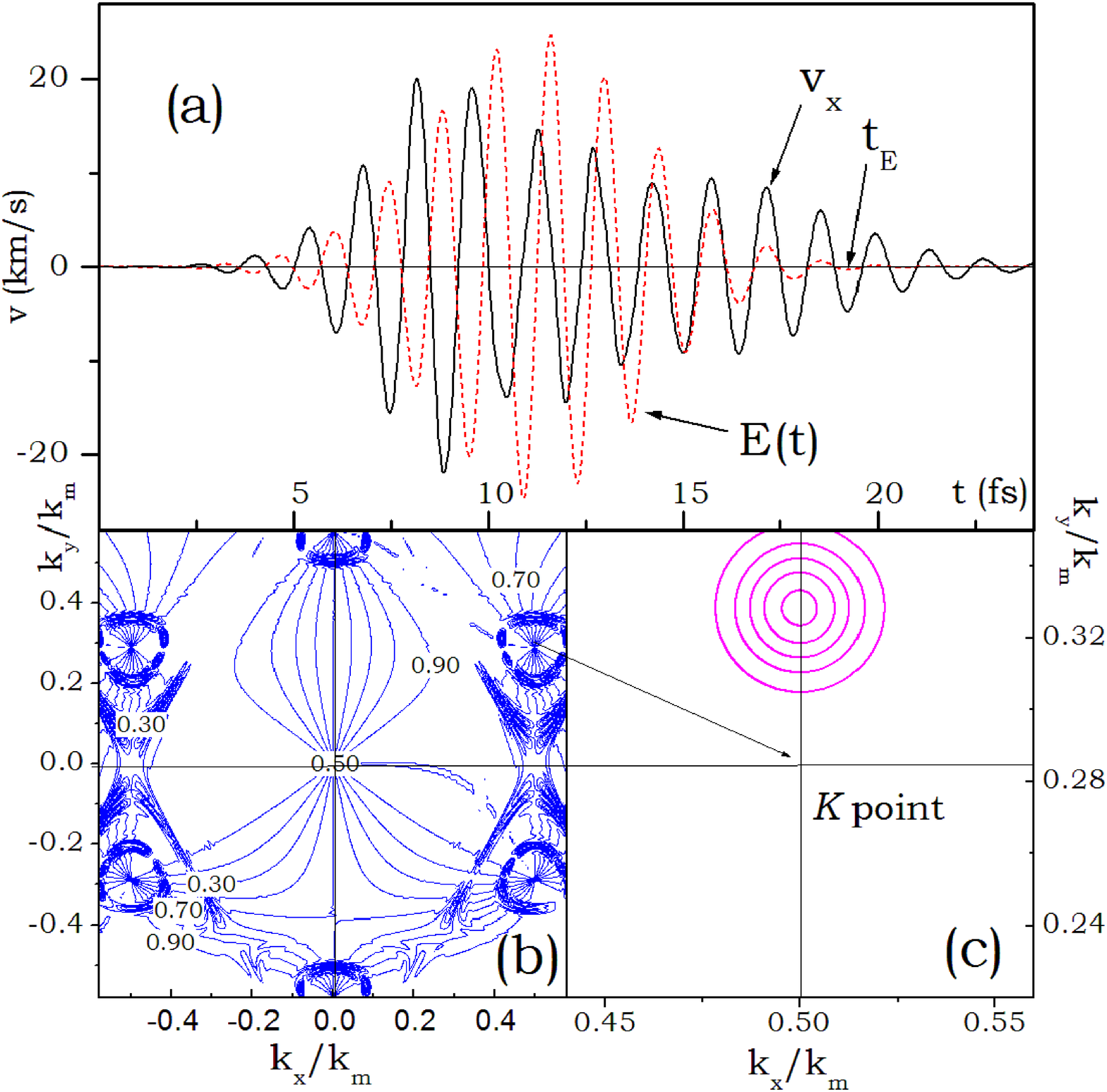}
\caption{(a) Average electron packet velocity, as calculated for monolayer graphene with the use of wave
             packet created by the lase pulse. Material and laser parameters are listed in table~1.
             Solid line: packet velocity, dashed line: electric field of the laser pulse (in arbitrary units);
         (b) Probability density~$P^-_P({\bm k})$ for~$t=t_E$ for the packet created by laser pulse;
         (c) Probability density~$P^-_G({\bm k})$ for~$t=0$ for the Gaussian packet given in
             equation~(\ref{Intr_Gauss}) and with parameters used in~\cite{Rusin2007b}.
             The arrow indicates position of one of the~$K$ points in the Brillouin zone of graphene.
            } \label{Fig8}
\end{figure}

The results shown in figures~5(c) and~6(c) for CNT indicate that the linear energy bands with~$m=6$ and~$\epsilon=\pm 1$
do not contribute to the ZB motion after the termination of laser pulse.
This suggests that the ZB oscillations may quickly disappear also in monolayer graphene, in which
the linear bands are most important.
To verify this conjecture we calculate the electron motion in graphene after an illumination by the laser pulse.
Introducing the vector potential into the Hamiltonian~(\ref{ML_hH}) in the way described in Sec.~2 we obtain
the following Schrodinger equation, see~(\ref{ML_hH}),~(\ref{ML_HAB2}) and~(\ref{CNT_h0})
\begin{equation} \label{AppB_hHt}
i\hbar\frac{d}{dt}\left(\begin{array}{c} \Psi_1(t) \\ \Psi_2(t)\end{array} \right) =
 t_{AB}\left(\begin{array}{cc} 0 & {\cal H}_{AB}^*(t) \\ {\cal H}_{AB}(t) & 0 \end{array} \right)
\left(\begin{array}{c} \Psi_1(t) \\ \Psi_2(t)\end{array} \right),
\end{equation}
where
\begin{equation} \label{AppB_HAB2t}
 {\cal H}_{AB}(t)= e^{iq_x(t)a/\sqrt{3}} + 2e^{-iq_x(t)a/(2\sqrt{3})}\cos\left(\frac{q_y(t)a}{2}\right),
\end{equation}
in which~${\bm q}(t) = {\bm k} -(e/\hbar){\bm A}(t)$. Then the average packet velocity is,
see~(\ref{CNT_vxkxmy}) and~(\ref{CNT_vx})
\begin{equation} \label{AppB_ML_vx}
 \langle v_x(t)\rangle=\frac{1}{(2\pi)^2}\int_{BZ} \left\langle \Psi_{\bm k}(t)\left|\frac{\partial \hat{H}(t)}{\hbar\partial k_x}
                       \right|\Psi_{\bm k}(t)\right\rangle d^2{\bm k},
\end{equation}
The integration is performed over the two-dimensional Brillouin zone (BZ),
which is a hexagon having vertices at the~$K$ and~$K'$ points:~$(2\pi/a)(0, \pm 2/3)$
and~$(2\pi/a)(\pm 1/\sqrt{3}, \pm 1/3)$. The average electron velocity in the~$y$ direction is obtained by analogy.
We solve numerically the Schrodinger equation~(\ref{AppB_hHt}) for~$(2N_x+1)\times (2N_y+1)$
values of~$k_x$ and~$k_y$, respectively, taking~$N_x=N_y=120$.
The initial condition for the wave function is:~$\Psi(0)=w^-$, where
\begin{equation} \label{AppB_ML_wm}
 w^- = \frac{1}{\sqrt{2}|H_{AB}|} \left(\begin{array}{c} -H_{AB}^* \\ |H_{AB}|\end{array} \right),
\end{equation}
is the negative energy eigenstate of the Hamiltonian~(\ref{ML_hH}).
The results of calculations for the material and pulse parameters listed in table~1 are shown in figure~8(a).
The solid line indicates~$\langle v_x(t)\rangle$ obtained from~(\ref{AppB_ML_vx}),
the dashed line is the laser electric field in arbitrary units. Average packet velocity in the~$y$ direction is zero.
As before,~$t_E\simeq 19$~fs indicates the time for which the pulse vanishes.
It is seen in figure~8(a) that the velocity oscillations disappear quickly after the pulse termination.
This result confirms our expectation that the laser pulse allows one to excite the ZB oscillations in nanotubes, but it
can hardly create the ZB motion in graphene.

Next we calculate the Fourier transform of~$\langle v_x(t)\rangle$ and find that the spectrum~$|\langle v_x(\omega)\rangle|^2$
contains the laser frequency~$\omega_L=4.5$~s$^{-1}$ plus two satellites at~$\omega_1\simeq 3.8$~s$^{-1}$
and~$\omega_2\simeq 5.1$~s$^{-1}$. This means that the electron oscillates with the laser frequency
and not with the interband frequency. This is in contrast to nanotubes
in which the electron, even during the pulse, oscillates
with~$\omega_Z$ corresponding to the interband energy between some pair of energy bands.
To confirm this conclusion we calculate the average packet velocity and
its Fourier transform for the pulse created by a Ti:Sapphire laser:~$\omega_L=2.4$~s$^{-1}$
and~$\tau=6.5$~fs. In this case the electron motion also disappears quickly after the pulse termination and its
Fourier spectrum contains the frequency~$\omega=\omega_L$ plus one satellite. These results confirm that
it is practically impossible to excite the ZB oscillations in monolayer graphene by a laser pulse.

The above conclusion seems to be inconsistent with our previous results obtained in~\cite{Rusin2007b,Rusin2013a},
in which the ZB oscillations in monolayer graphene were calculated for Gaussian packets.
This apparent contradiction is caused by the difference
between electron wave packets used in the two cases. To show this difference we again calculate probability densities
of states having negative energies for both packets, in a way analogous to that described in Section~2,
see~(\ref{CNT_a1a2}) and the last paragraph of Section~2.
In figure~8(b) we plot the probability density~$P^-_P({\bm k})$ at~$t=t_E$ for the wave packet created by laser pulse.
The magnitude of the wave vector is expressed in~$k_m=4\pi/\sqrt{3}a=2.95\times 10^{10}$~m$^{-1}$ units.
The wave packet is calculated numerically by solving the Schrodinger equation~(\ref{AppB_hHt}).
The packet shown in figure~8(b) is delocalized in the~${\bm k}$ space and it consists
of wave vectors within the whole BZ. It is seen that there are contributions to~$P^-_P({\bm k})$ arising
from the six vertices of the hexagonal BZ of graphene and from other areas of the BZ, e.g. from its center.
The probability distribution does not possess the sixfold symmetry because the laser light is assumed to be
polarized in the~$x$ direction. On the other hand,
in figure~8(c) we plot the probability density~$P^-_G({\bm k})$ for the Gaussian packet given in~(\ref{Intr_Gauss}),
taking~$k_{0x}=0$,~$k_{0y}=1.2\times 10^9$ m$^{-1}$ and~$d=20$~\AA, see~\cite{Rusin2007b}.
In our previous approach the Hamiltonian~(\ref{ML_hH}) was expanded in the frame of~${\bm k} \cdot {\bm p}$ theory
in the vicinity of the~$K$ point of BZ. The arrow indicates position of one the~$K$ points in BZ.
Note the big difference in scales in figure~8(b) and figure~8(c).
In contrast to the distribution shown in figure~8(b), the Gaussian packet in figure~8(c) is concentrated in a small fraction
of BZ. For this reason, one may expect very different behavior of ZB oscillations for both packets,
which is indeed the case, see figure~8(a) and~\cite{Rusin2013a}. We conclude that the results obtained in this work for graphene are
not inconsistent with the results obtained in~\cite{Rusin2013a}.

\section{Vector potential}

\begin{figure}
\includegraphics[width=8.0cm,height=8.0cm]{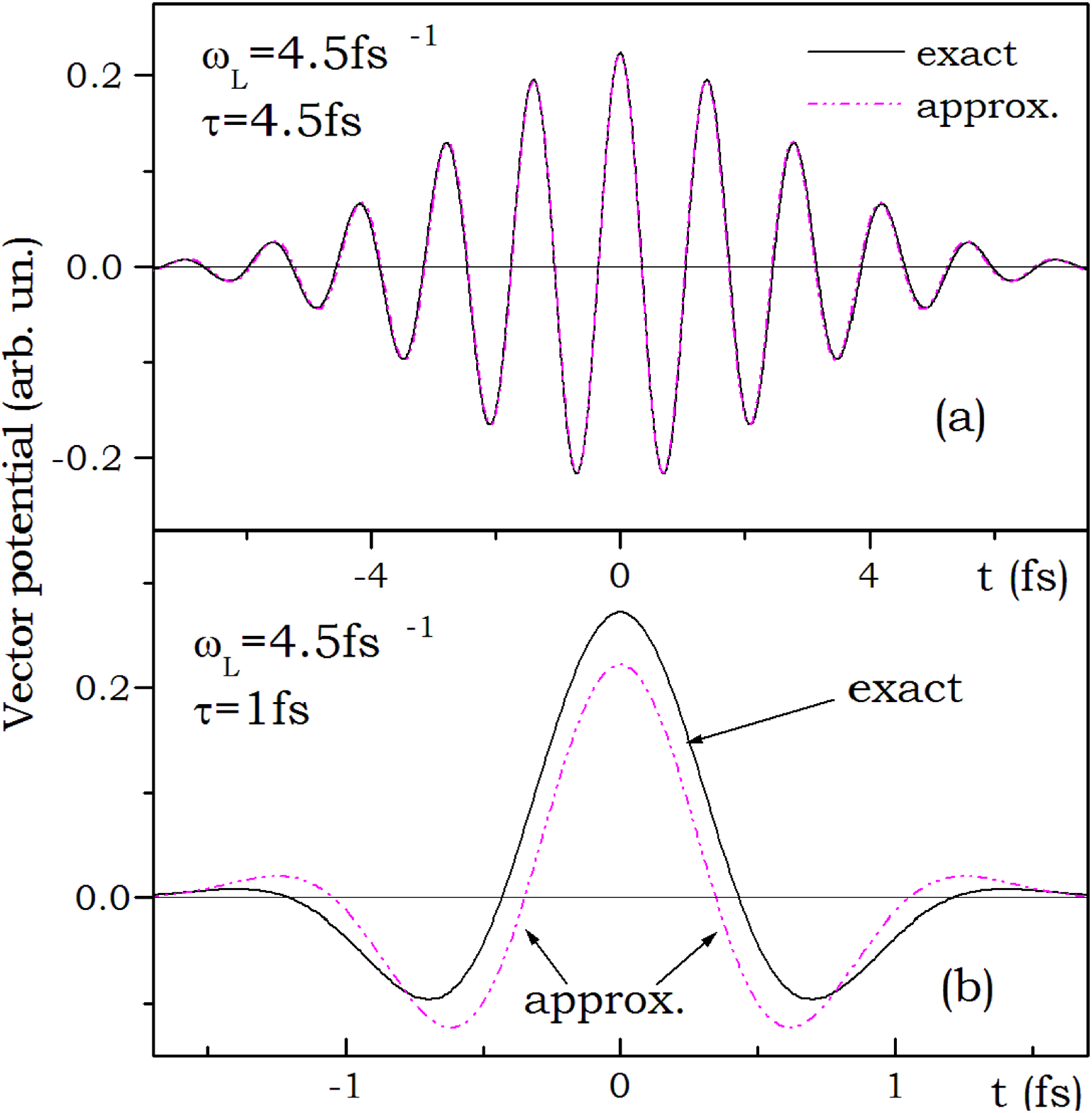}
\caption{Vector potential versus time calculated exactly (solid lines) and with the use of approximate formula
         (\ref{Pulse_At}) (dash-dotted lines) for two sets of pulse parameters. For the upper panel the approximate results are
         practically indistinguishable from exact ones.} \label{Fig9}
\end{figure}
Here we discuss an approximate form of the vector potential, as given in~(\ref{Pulse_At}),
which is frequently used in the physics of short laser pulses, see e.g.~\cite{Bandrauk2002}.
The electric field~$E(t)$ in~(\ref{Pulse_Et})
is not obtained exactly from the vector potential~(\ref{Pulse_At}).
There is:~$-\partial A(t)/\partial t = E(t) + \delta E$,
where~$\delta E$ is very small for~$\tau$ and~$\omega_L$ listed in
table~1, so it can be neglected. To show the validity of this approximation we show in figure~9 the
exact vector potential:~$ A^{ex}(t) = -\int_{-\infty}^{t} E(t') dt'$ (solid lines) and the approximate one given
in~(\ref{Pulse_At}) (dash-dotted lines). For pulse parameters listen in table~1 (upper panel) the two
curves are indistinguishable, which justifies approximate form of the vector potential, as used in our calculations.
For nearly monocycle pulses (lower panel) with~$\tau=1$~fs the approximation is
still qualitatively correct, but small deviations from the exact values are visible.
Our approximation fails only for sub-monocycle pulses.
We note that one could plot figures~3,~5 and~6 as functions of the vector
potential~$A(t)$ rather than in terms of electric field.

---------------------------------------------------------------

\end{document}